\documentclass[10pt,english,aps,prl,superscriptaddress,nofootinbib,notitlepage,preprintnumbers,twocolumn]{revtex4-2}
\usepackage[justification=raggedright]{caption}
\usepackage{booktabs}
\usepackage[T1]{fontenc}
\usepackage[latin9]{inputenc}
\usepackage{color}
\usepackage{bm}
\usepackage{esint}
\usepackage{graphicx}
\usepackage{comment}
\graphicspath{{images/}}
\usepackage[caption=false]{subfig}
\makeatletter
\AtBeginDocument{\usepackage{booktabs}}
\def\Mpl{M_{\rm P}}

\usepackage{babel}
\makeatother
\begin{document}
\preprint{YITP-20-117, IPMU 20-0098}
\title{Addressing $H_0$ tension by means of VCDM}

\author{Antonio De~Felice}
\email{antonio.defelice@yukawa.kyoto-u.ac.jp}

\affiliation{Center for Gravitational Physics, Yukawa Institute for Theoretical
Physics, Kyoto University, 606-8502, Kyoto, Japan}

\author{Shinji Mukohyama}
\email{shinji.mukohyama@yukawa.kyoto-u.ac.jp}

\affiliation{Center for Gravitational Physics, Yukawa Institute for Theoretical
Physics, Kyoto University, 606-8502, Kyoto, Japan}
\affiliation{Kavli Institute for the Physics and Mathematics of the Universe (WPI),
The University of Tokyo, Kashiwa, Chiba 277-8583, Japan}

\author{Masroor C.~Pookkillath}
\email{masroor.cp@yukawa.kyoto-u.ac.jp}

\affiliation{Center for Gravitational Physics, Yukawa Institute for Theoretical
Physics, Kyoto University, 606-8502, Kyoto, Japan}

\begin{abstract}
  In this letter we propose a reduction of the $H_0$ tension puzzle by
  means of a theory of minimally modified gravity which is dubbed
  VCDM. After confronting the theory with the experiments, we find that the
  data allow for a low-redshift transition in the expansion history of the
  universe at either $z\simeq 0.3 $ or $z \simeq 1.8\,$, corresponding to one
  of the two local minima of the total $\chi^2$. From the bestfit
  values the total fitness parameter is improved by
  $\Delta \chi^2 \simeq 12$, for the data set considered. We then infer the
  local Hubble expansion rate today within this theory by means of low
  redshift Pantheon data. The resulting local Hubble expansion rate
  today is $H^{\rm{loc}}_0=73.6\pm1.4$. We find the tension is reduced
  within the VCDM theory.
\end{abstract}

\maketitle


The value of today's rate of expansion of the universe, $H_0$, has
been measured, as a direct measurement, from low-redshift
observations, namely, SH$0$ES~\cite{Riess:2019cxk},
H0LiCOW~\cite{Wong:2019kwg}, Megamaser Cosmology Project~\cite{Reid:2008nm}
(MCP) and Carnegie-Chicago Hubble Program (CCHP)
Collaboration~\cite{Freedman:2019jwv}. Among these observations
SH$0$ES in particular has achieved a remarkable precision providing
$H_0=\textrm{74.03}\pm\textrm{1.42}$ (in units of $\rm{km\, s^{-1}\,
  Mpc^{-1}}$). On the other hand, on assuming some theoretical model,
$H_0$ can also be deduced from the measurement of temperature power
spectra in the Cosmic Microwave Background (CMB) which is
produced at the recombination time. The recent Planck Legacy 2018
release gives $H_0=\textrm{67.04}\pm\textrm{0.5}$, assuming the
standard flat-$\Lambda$CDM model (non-flat versions are known to be
strongly disfavored by other data, e.g.\ Baryon Acoustic Oscillations
(BAO))~\cite{Aghanim:2018eyx}. Hence, the tension between this
theoretical model and experimental results adds up to more than
4$\sigma$'s~\cite{Bernal:2016gxb,Riess:2020sih}.

However, the flat-$\Lambda$CDM could be representing only a first
approximation to another, improved model of our universe. The VCDM
theory, described in the following, was originally introduced for the
purpose of seeking minimal theoretical deviations from the standard
model of gravity and cosmology, i.e.\ General Relativity (GR) and
$\Lambda$CDM, as it does not introduce any new propagating physical
degrees of freedom in the gravity sector, but on the other hand, one
can have, as we will show in the following, a non-trivial and
interesting phenomenology.


In the VCDM theory~\cite{DeFelice:2020eju}, the cosmological
constant $\Lambda$ in the standard $\Lambda$CDM is promoted to a
function $V(\phi)$ of a non-dynamical, auxiliary field $\phi$.
This theory of modified
gravity breaks four dimensional diffeomorphism invariance at
cosmological scales but keeps the three dimensional spatial
diffeomorphism invariance. On doing so, the theory modifies gravity at
cosmological scales while it only possess two gravitational degrees of
freedom as in GR. In general this allows a spectrum of possibilities
typically much larger than the case of a scalar-tensor theory. For the
latter, the extra scalar degree of freedom leads to strong constraints not only 
on solar system scales (for which one needs the scalar to be very
massive or to be shielded by some non-trivial dynamical mechanisms,
e.g.\ chameleon or Vainshtein), but also on cosmological scales (for
which one needs to constrain the background dynamics as to avoid ghost
and gradient instabilities).

The equations of motion for the VCDM theory on a homogeneous and
isotropic background can be written as
\begin{equation}
V = \frac{1}{3}\phi^2 -\frac{\rho}{\Mpl^2}\,, \quad
\frac{d\phi}{d\mathcal{N}} = \frac{3}{2}\frac{\rho+P}{\Mpl^2H}\,,\quad
\frac{d\rho_I}{d\mathcal{N}} +3(\rho_I +
P_I)=0\,, \label{eqn:backgroundEOM}
\end{equation}
where $\mathcal{N}=\ln(a/a_0)$ ($a$ being the scale factor and $a_0$
being its present value), $H=\dot{a}/a^2$ is the Hubble expansion rate
(a dot denotes differentiation with respect to the conformal time),
$\rho=\sum_I\rho_I$ and $P=\sum_IP_I$ (the sum is over the standard
matter species). Unless $\rho+P=0$, the following equation follows
from the above equations: $\phi = \frac{3}{2}V_{,\phi} - 3H$. When $V$
is a linear function of $\phi$, as in $V=\lambda_1 \phi+\lambda_0$,
then the equations of motion (\ref{eqn:backgroundEOM}) reduce to
\begin{equation}
 3H^2 = \frac{\rho}{\Mpl^2}+\Lambda\,, \quad H\frac{dH}{d\mathcal{N}}
 = -\frac{\rho+P}{2\Mpl^2}\,, \quad \frac{d\rho_i}{d\mathcal{N}}
 +3(\rho_I + P_I)=0\,,
\end{equation}
where $\Lambda\equiv \lambda_0+3\lambda_1^2/4={\rm const}$. These are
nothing but the equations of motion in the standard $\Lambda$CDM
model. Moreover, for this choice of $V(\phi)$ the theory reduces to GR with 
a cosmological constant not only for the homogeneous and isotropic background 
but also for perturbations at any order. 
Hence, the VCDM theory extends the $\Lambda$CDM model by
replacing the constant $\Lambda$ with a free function $V(\phi)$. Yet,
the VCDM theory does not introduce extra degrees of freedom in the
sense that the number of independent initial conditions is the same as
$\Lambda$CDM. The ``V'' of VCDM therefore stands for the free function
$V(\phi)$ introduced in this theory.

In the following we want to be able to use the free function $V(\phi)$
in order to give any wanted background evolution for $H$ which can be
given as $H=H(\mathcal{N})$. From the 2nd of
(\ref{eqn:backgroundEOM}), having given $H$ as a function of
$\mathcal{N}$, then one obtains
\begin{equation}
 \phi(\mathcal{N}) = \phi_0 +
 \int_{\mathcal{N}_0}^{\mathcal{N}}\frac{3}{2}\frac{\rho(\mathcal{N}')+P(\mathcal{N}')}{M_{\rm
     Pl}^2H(\mathcal{N}')}d\mathcal{N}'\,, \label{eqn:phi}
\end{equation}
where $\phi_0=\phi(\mathcal{N}_0)$. Assuming that $\rho + P > 0$, and
$H > 0$, the right hand side of (\ref{eqn:phi}) is an increasing
function of $\mathcal{N}$ and thus the function $\phi(\mathcal{N})$
has a unique inverse function, $\mathcal{N} =
\mathcal{N}(\phi)$. Obviously, $\mathcal{N}$ is an increasing function
of $\phi$. By combining this with the 1st of
(\ref{eqn:backgroundEOM}), one obtains
\begin{equation}
 V(\phi) = \frac{1}{3}\phi^2 - \frac{\rho(\mathcal{N}(\phi))}{M_{\rm
     Pl}^2}\,.
\end{equation}
Therefore we have obtained a simple and powerful reconstruction
mechanism for $V$ for a given and wanted evolution of
$H$~\footnote{The potential $V(\phi)$ is not
    unique. Even for the same choice of $H(z)$, the shape of the potential $V(\phi)$ depends on the initial value of $\phi$, which can be chosen arbitrarily. Changing the initial value of $\phi$ changes the potential $V$ by a constant and a term linear in $\phi$. In any case, the potential does not enter both at the level of the background and linear perturbation theory which are affected only by $H(z)$ and its derivatives. Furthermore, VCDM theory reduces to GR whenever $V_{,\phi\phi}=0$. See e.g. eq.~(5.11) of \cite{DeFelice:2020eju}, which shows $w_{\phi}=-1$ whenever $V_{,\phi\phi}=0$}. Once $V(\phi)$ is specified in this
way, we know how to evolve not only the homogeneous and isotropic
background but also perturbations around it.

Having shown the reconstruction of the potential $V$ for a given
evolution of $H$, we can search for a profile for $H(z)$ that can potentially
reduce the $H_{0}$ tension. We introduce two
choices of $H(z)$ to address current cosmological tensions: one in the
flat-$\Lambda$CDM and the other in the VCDM. The former is
$H_{\Lambda}^{2}\equiv
H_{\Lambda0}^{2}[\tilde{\Omega}_{\Lambda}+\tilde{\Omega}_{m0}(1+z)^{3}+\tilde{\Omega}_{r0}(1+z)^{4}]$,
where
$\tilde{\Omega}_{\Lambda}\equiv1-\tilde{\Omega}_{m0}-\tilde{\Omega}_{r0}$,
and the latter is
\begin{equation}
H^{2}=H_{\Lambda}^{2}+A_{1}H_{0}^{2}\left[1-\tanh\left(\frac{z-A_{2}}{A_{3}}\right)\right],
\end{equation}
with the idea that $0<A_{2}<2$. In this case we have at early times,
for $z\gg |A_{2}|$, that the system will tend to be the standard
flat-$\Lambda$CDM evolution, i.e.\ $H^{2}\approx
H_{\Lambda}^{2}$. Today, i.e. for $z=0$, we have
$H_{0}^{2}=H_{\Lambda}^{2}+A_{1}H_{0}^{2}\left[1+\tanh\left(\frac{A_{2}}{A_{3}}\right)\right]$,
which can be solved today for $A_{1}$, to give
$A_{1}=(1-H_{\Lambda0}^{2}/H_{0}^{2})\left[\tanh\left(A_{2}/A_{3}\right)+1\right]^{-1}$.

Let us further consider the following parameter redefinitions
$\tilde{\Omega}_{m0} = \Omega_{m0}\,H_{0}^{2}/H_{\Lambda0}^{2}$,
$\tilde{\Omega}_{r0} = \Omega_{r0}\,H_{0}^{2}/H_{\Lambda0}^{2}$, and
$\beta_{H} = H_{\Lambda0}/H_{0}$, then we find\footnote{As for Montecarlo sampling, to make sure about convergence, we have also run the chains by redefining
  the parameter $\beta_H$ as $\beta_H =(2-{\tilde\beta}_H^2)^{-1/2}$ and have given a flat prior for the new parameter ${\tilde\beta}_H$.}
\begin{widetext}
\begin{equation}
  \frac{H^{2}}{H_0^2} = \Omega_{{\rm m}0}(1+z)^{3} + \Omega_{{\rm
      r}0}(1+z)^{4} +
  (1-\beta_{H}^{2})\frac{1+\tanh\!\left(\frac{A_{2}-z}{A_{3}}\right)}{1+\tanh\!\left(\frac{A_{2}}{A_{3}}\right)}
  + \beta_{H}^{2}\left(1-\frac{\Omega_{{\rm
        m}0}}{\beta_{H}^{2}}-\frac{\Omega_{{\rm
        r}0}}{\beta_{H}^{2}}\right) \,.
  \label{eq:Fried}
\end{equation}
\end{widetext}
So in total we have six background parameters (three more than
$\Lambda$CDM). However, we can reduce them to five (two more than
$\Lambda$CDM) by fixing $A_{3}$ as we expect to have a large
degeneracy (after assuming $A_2 = \mathcal{O}(1)$). According to
Akaike Information Criterion (AIC), we can accept the model if we can
have an improvement of $\chi^2$ larger than four in comparison with
$\Lambda$CDM~\cite{Liddle:2007fy}. In fact, we will show later on that
the $\chi^2$ has improved by $\Delta \chi^2 \simeq 12$ with respect to
$\Lambda$CDM. In particular, we will fix, later on, $A_3$ to the value
of 10${}^{-3}$.

Two things need to be noticed. First, having given the expression for
$H=H(z)$, one can automatically deduce all the needed background
expressions as well as all evolution equations for perturbations. 
Second, the fact we have a minimally modified gravity
(VCDM)-component does not mean we are adding a physical dark-component
degree of freedom. In fact, for this theory, there is no additional
physical degree of freedom, beside the tensorial gravitational waves
and the standard ones related to the presence of matter
fields~\cite{DeFelice:2020eju}.


After having introduced the behavior of the VCDM model on a
homogeneous and isotropic background, we will test it against several
cosmological data to see how well it can address the $H_0$
tension. Here we use Planck Legacy 2018 data with
\texttt{planck\_highl\_TTTEEE}, \texttt{planck\_lowl\_EE}, and
\texttt{planck\_lowl\_TT}~\cite{Aghanim:2019ame}, baryon acoustic
oscillation (BAO) from 6dF Galaxy Survey~\cite{Beutler:2011hx} and the
Sloan Digital Sky Survey~\cite{Ross:2014qpa,Alam:2016hwk}, and
Pantheon data set comprised of 1048 type Ia
supernovae~\cite{Scolnic:2017caz}. We use single data points of $H_0$
from SH0ES $H_0=73.2\pm1.3$~\cite{Riess:2020fzl} (in standard units),
H0LiCOW $H_0=73.3\pm1.8$~\cite{Wong:2019kwg} and MCP
$H_0=73.9\pm3$~\cite{Reid:2008nm}. We also add to the Pantheon
likelihood a gaussian likelihood for the absolute magnitude $M$ of the
supernovae samples to properly account astrophysical independent input
inside the context of the Hubble
tension~\cite{Camarena:2019moy,Benevento:2020fev,Camarena:2021jlr,Efstathiou:2021ocp}. We
then infer the local value of Hubble expansion today $H^{\rm{loc}}_0$
using the Pantheon data and employing the analysis technique explained
in~\cite{Camarena:2019moy,Benevento:2020fev}. Here we have not
included the Planck lensing data since it was reported that the
lensing anomaly is present in the Plank legacy
release~\cite{Aghanim:2016sns,Motloch:2019gux}.

Both the background and linear perturbation equations of motion are
implemented in the Boltzmann code CLASS~\cite{Blas:2011rf}, with
covariantly corrected baryon equations of
motion~\cite{Pookkillath:2019nkn}.

For a matter action at second order up to shear for a fluid we proceed
here by first writing the Schutz-Sorkin Lagrangian (SSL) for a single
fluid~\cite{Pookkillath:2019nkn}, as follows
\begin{widetext}
\begin{equation}
S_{{\rm SSL}}=-\int
d^{4}x\sqrt{-g}[\rho(n,s)+J^{\mu}\,(\partial_{\mu}\ell+\theta\partial_{\mu}s+A_{1}\partial_{\mu}B_{1}+A_{2}\partial_{\mu}B_{2})]\,,
\end{equation}
\end{widetext}
with $n=\sqrt{-J^{\mu}J^{\mu}g_{\mu\nu}}$, and 4-velocity
$u^\alpha=J^\alpha/n$.  We consider several copies of the previous
action each describing a different fluid, labeled with an index
$I$. Then we can expand the previous SSL up to second order in the
perturbation fields, and to this one, we then add a correction aimed
to describe an anisotropic fluid as follows
$S_{{\rm m}}^{(2)}=S_{{\rm SSL}}^{(2)}+S^{(2)}{}_{{\rm corr}}$, where
$S^{(2)}{}_{{\rm corr}}=\int dtd^{3}xNa^{3}\sum_{I}\sigma_{I}\,\Theta_I$
and $\Theta_I$ stands for a linear combination of perturbation
fields. Since for each matter species $\rho_I=\rho_I(n_I)$, and
$n_I=\sqrt{-J_I^{\mu}J_I^{\nu}g_{\mu\nu}}$, one can find a relation
among $\delta\rho_I$ and the other fields as
$\delta
J_{I}=\frac{\rho_{I}}{n_{I}\rho_{I.n}}\,\frac{\delta\rho_{I}}{\rho_{I}}-\alpha$,
where $\delta N = N(t)\alpha$, which can be used as a field
redefinition to replace $\delta J_{I}$ in terms of
$\frac{\delta\rho_{I}}{\rho_{I}}$. We also define gauge invariant
combinations
$v_{I} =
-\frac{a}{k^{2}}\,\theta_I+\chi-\frac{a^2}{N}\partial_t(E/a^2)$,
$\alpha =
\Psi-\frac{\dot{\chi}}{N}+N^{-1}\partial_t[a^2N^{-1}\partial_t(E/a^2)]$,
and
$\zeta =
-\Phi-H\,\chi+\frac{a^{\mathrm{2}}H}{N}\,\partial_t(E/a^{2})$, where
$\delta\gamma_{ij}=2[a^2\zeta\delta_{ij}+\partial_i\partial_j E]$, and
$\delta u_{Ii}=\partial_i v_I$.  We find finally that
\begin{equation}
S_{{\rm corr}}^{(2)} = \int
dtd^{3}x\,Na^{3}\sum_{I}\sigma_{I}\left[\delta\rho_{I}+3\,(\rho_{I}+P_{I})\,\zeta\right].
\end{equation}

For vector transverse modes, we can define $ T_{I}^{i}{}_{j} \equiv
P_{I}\,\delta^{i}{}_{j}+P_{I}\,\frac{\delta^{ik}}{a^{2}}\,\pi_{kj}^{I}$, and $\pi_{ij}^{I} \equiv
\frac{1}{2}\,(\partial_{i}\pi_{j}^{I,T}+\partial_{j}\pi_{i}^{I,T})$. Then we can introduce the 1+3 decompositions for the 4-velocity of the fluid $u_{Ii}^{V,I}  = \delta u_{Ii}^{I}$, the shift $N_{i} =a\,N\,G_{i}$, and the 3D metric $\delta\gamma_{ij} = a\,(\partial_{i}C_{j}+\partial_{j}C_{i})$. We can also introduce the following gauge invariant variables $V_{i}=G_{i}-\frac{a}{N}\frac{d}{dt}\!\left(\frac{C_{i}}{a}\right)$, and $F_{i}^{I} = \frac{C_{i}}{a}-\frac{b_{\mathrm{1}i}^{I}}{\vec{b}_{\mathrm{1}}^{I}\cdot\vec{b}_{\mathrm{1}}^{I}}\,\delta B_{\mathrm{1}}^{I}-\frac{b_{\mathrm{2}i}^{I}}{\vec{b}_{\mathrm{2}}^{I}\cdot\vec{b}_{\mathrm{2}}^{I}}\,\delta B_{\mathrm{2}}^{I}$, where $b_{\mathrm{1}i}^{I} b_{\mathrm{2}i}^{I}=0=b_{\mathrm{1}i}^{I}k^i=b_{\mathrm{2}i}^{I}k^i$. Then, 
on following a similar approach one finds
the total Lagrangian density for the vector
perturbations, in VCDM, can be written as
\begin{widetext}
\begin{eqnarray}
\mathcal{L} & = & Na^{3}\delta^{ij}\left\{
\sum_{I}n_{I}\rho_{I,n}\,\frac{\dot{F}_{i}^{I}}{N}\,\delta
u_{j}^{I}+\frac{1}{a^{2}}\sum_{I}n_{I}\rho_{I,n}\left[a\delta
  u_{i}^{I}V_{j}-\frac{1}{2}\,\delta u_{i}^{I}\delta
  u_{j}^{I}\right]-\frac{\Mpl^2}{4a^{2}}\,V_{i}(\delta^{lm}\partial_{l}\partial_{m}V_{j})\right.\nonumber
\\ & &
   {}-\left.\frac{1}{2a^{2}}\sum_{I}P_{I}\,\pi_{i}^{I,T}\,(\delta^{lm}\partial_{l}\partial_{m}F_{j}^{I})\right\}
   ,
\end{eqnarray}
\end{widetext}
which reduces to the same result as in GR. 

Finally, for the tensor modes, let us define $\delta \gamma_{ij}  =a^{2}\,(h_{+}\varepsilon_{ij}^{+}+h_{\times}\varepsilon_{ij}^{\times})$, where $\varepsilon_{ij}^{+,\times}   =  \varepsilon_{ji}^{+,\times}$, $\delta^{ij}\varepsilon_{ij}^{+,\times}  = 0$, $\varepsilon_{ij}^{+}\varepsilon_{mn}^{\times}\delta^{im}\delta^{jn}=0$, and $\varepsilon_{ij}^{+}\varepsilon_{mn}^{+}\delta^{im}\delta^{jn} =   1=  \varepsilon_{ij}^{\times}\varepsilon_{mn}^{\times}\delta^{im}\delta^{jn}$. As for the energy-momentum tensor we have instead for the perturbations $\delta T_{I}^{i}{}_{j}  \equiv P_{I}\,\frac{\delta^{ik}}{a^{2}}\,\pi_{kj}^{I,TT}$, so that the total Lagrangian density in VCDM becomes
\begin{widetext}
\begin{equation}
\mathcal{L}=\frac{\Mpl^{2}}{8}\,\frac{a^{3}}{N}\,(\dot{h}_{+}^{2}+\dot{h}_{\times}^{2})-\frac{Na\,\Mpl^{2}}{8}\,[(\partial_{i}h_{+})\delta^{ij}(\partial_{j}h_{+})+(\partial_{i}h_{\times})\delta^{ij}(\partial_{j}h_{\times})]+\frac{Na}{2}\sum_{I}P_{I}\,(h_{+}\pi_{+}^{I}+h_{\times}\pi_{\times}^{I})\,,
\end{equation}
\end{widetext}
which reduces to the same form of GR.

Before substituting the explicit dependence of $H$ on the redshift $z$, 
all the equations of motion (including the ones for the matter fields)
for the perturbations are, in form, exactly the same as for
$\Lambda$CDM, except the following one (written in terms
of the Newtonian-gauge invariant fields $\Phi$ and $\Psi$):
\begin{equation}
\dot{\Phi}+aH\Psi  = 
\frac{3\,[k^{2}-3a^{2}(\dot{H}/a)]}{k^{2}\,[2k^{2}/a^{2}+9\sum_{K}(\varrho_{K}+p_{K})]}\,\sum_{I}(\varrho_{I}+p_{I})\,\theta_{I}\,, \label{eqn:dotphi}
\end{equation}
which is used to find the evolution of the curvature perturbation
$\Phi$, and where a dot represents a derivative with respect the
conformal time. Here we have used CLASS notation, namely $\varrho_I=\rho_I/(3M_{\rm{P}}^2)$, $p_I=P_I/(3M_{\rm{P}}^2)$.
At the level of linear perturbation, the deviation from $\Lambda$CDM therefore consists of two parts: the explicit difference seen in (\ref{eqn:dotphi}) and the implicit difference due to different $H(z)$.

The parameter estimation is made via Markov Chain Monte Carlo (MCMC)
sampling by using Monte Python~\cite{Audren:2012wb,Brinckmann:2018cvx}
against the above mentioned data sets. In the MCMC sampling we used
very high precision by decreasing the step size for both background
and perturbation integration\footnote{More in detail we have set in CLASS the variables \texttt{tol\_perturb\_inte\-gra\-tion, tol\_back\-ground\_in\-te\-gra\-tion} to $10^{-12}$, \texttt{back\_in\-te\-gra\-tion\_step\-size, perturb\_in\-te\-gra\-tion\_step\-size} to $10^{-4}$, etc.} to see the smooth transition of
$H(z)$ for both VCDM and $\Lambda$CDM. The analysis of the MCMC chains is performed using a chain
analyzer package, GetDist~\cite{Lewis:2019xzd}.

We have considered the prior for the parameters of VCDM such that
$\Lambda$CDM is well inside the region. (We will see later on that
there are two local minima of the total $\chi^2$, corresponding to 
transitions at redshift $z \simeq 0.3$
(low-$z$) and $z \simeq 1.8$ (high-$z$), respectively). In particular,
for low-$z$ we give the following priors:
$\textrm{0.6}<\beta_H<\textrm{1.3}$, and
$\textrm{0.15}<A_2<\textrm{0.4}$. For high-$z$ we give
$\textrm{0.05}<\beta_H<\textrm{2.3}$ ($\beta_H\to0$ just implies that
the amount of amplitude transition is finite), and
$\textrm{0.4}<A_2<\textrm{5}$. We fix $A_3=\textrm{10}^{-3}$ as it has
large degeneracy in both low-$z$ and high-$z$. Deviations of $\beta_H$
from 1 imply that the cosmological data sets prefer the VCDM model
over $\Lambda$CDM.


By doing the chain analysis we found that, on using only
Planck-BAO-Pantheon data alone, VCDM shows two minima: one has a value
of $\beta_H\simeq0.94$, and the other one has a lower value of $\beta_H$. The two minima
(for the $\chi^2$-function) differ, among other things, on the value for the
redshift at which the VCDM model shows the transition for $H(z)$.

In order to discriminate between these two minima, we also add to the
previous data sets three robust measurements of $H_0$, namely SH0ES,
H0LiCOW and MEGAMASER. On doing this, both minima still remain, but,
in any case, the total $\chi^2$ clearly favors VCDM, as $\beta_H<1$ at
2$\sigma$, whereas $\Lambda$CDM still gives a much lower value for
$h_0$ giving rise to the well-known tension among data sets.

One may wonder whether the data still point towards other possible
features which the chosen $H(z)$ (for this project) for VCDM might not
be able to fully implement. If in the near future more and more data
will be pointing towards some non-trivial features for $H(z)$, VCDM
(with practically for any chosen, but positive, $H(z)$), has to be
considered as a solid playground with which one can test modified
background cosmological evolutions.  In order to have a better picture
of the $\chi^2$ for the cosmological data sets we have considered, in
Table~\ref{chi2_eff-each}, we compare the contribution to the $\chi^2$
from each experiment between VCDM and $\Lambda$CDM and also show the
residue $\Delta \chi^2$.  Furthermore, since we find that at 2$\sigma$ $A_2>0.24$
($A_2$ fixes the redshift of the transition for $H(z)$ to a larger
value of $H$), the VCDM theory combined with the cosmological data
sets automatically avoids the potential problem of a transition around
$z=0$ as explained in \cite{Benevento:2020fev}.

\begin{table*}
\begin{tabular}{l l l l l}
\toprule
\textbf{Experiments} & \hspace{0.5cm}  \textbf{$\Lambda$CDM} & \hspace{0.5cm} \textbf{VCDM low$-z$ ($\Delta \chi^2$)} & \hspace{0.5cm} \textbf{VCDM high$-z$ ($\Delta \chi^2$)}\\
\midrule
Planck\_highl\_TTTEEE & \hspace{0.5cm} $2354.01$ & \hspace{0.5cm} $2349.56 \quad (4.45)$ & \hspace{0.5cm} $2347.03 \quad (6.98)$ \\
Planck\_lowl\_EE & \hspace{0.5cm} $397.37$ & \hspace{0.5cm} $395.92 \quad (1.45)$ & \hspace{0.5cm} $395.83 \quad (1.54)$ \\
Planck\_lowl\_TT & \hspace{0.5cm} $22.16$ & \hspace{0.5cm} $22.89 \quad (-0.73)$ & \hspace{0.5cm} $23.25 \quad (-1.09)$ \\
Pantheon & \hspace{0.5cm} $1027.28$  & \hspace{0.5cm} $1031.64 \quad (-4.36)$ & \hspace{0.5cm} $1027.31 \quad (-0.03)$ \\
bao\_boss\_dr12 & \hspace{0.5cm} $4.79$ & \hspace{0.5cm} $5.38 \quad (-0.59)$ & \hspace{0.5cm} $9.27 \quad (-4.48) $ \\
bao\_smallz\_2014 & \hspace{0.5cm} $3.14$ & \hspace{0.5cm} $5.31 \quad (-2.17)$ & \hspace{0.5cm} $4.58 \quad (-1.44)$ \\
absolute\_M & \hspace{0.5cm} $11.47$ & \hspace{0.5cm} $6.57 \quad (4.9)$ & \hspace{0.5cm} $6.85 \quad (4.62)$ \\  
  $H_0$ (SH0ES) & \hspace{0.5cm} $8.54$ & \hspace{0.5cm} $3.31 \quad (5.23)$ & \hspace{0.5cm} $4.34$ \quad (4.2)\\
$H_0$ (H0LiCOW) & \hspace{0.5cm} $4.69$ & \hspace{0.5cm} $1.88 \quad (2.81)$ & \hspace{0.5cm} $2.43 \quad (2.26)$ \\
$H_0$ (MEGAMASER) & \hspace{0.5cm} $2.25$ & \hspace{0.5cm} $1.04 \quad (1.21)$ & \hspace{0.5cm} $1.29 \quad (0.96)$ \\
\midrule
  Total & \hspace{0.5cm} $3835.71$ & \hspace{0.5cm} $3823.50 \quad (12.21)$ & \hspace{0.5cm} $3822.19 \quad (12.51)$\\
\bottomrule
\end{tabular}
\caption{Comparison of effective $\chi^2$ between VCDM and $\Lambda$CDM for individual data sets.}\label{chi2_eff-each}
\end{table*}

\begin{figure*}
\includegraphics[width=1\linewidth]{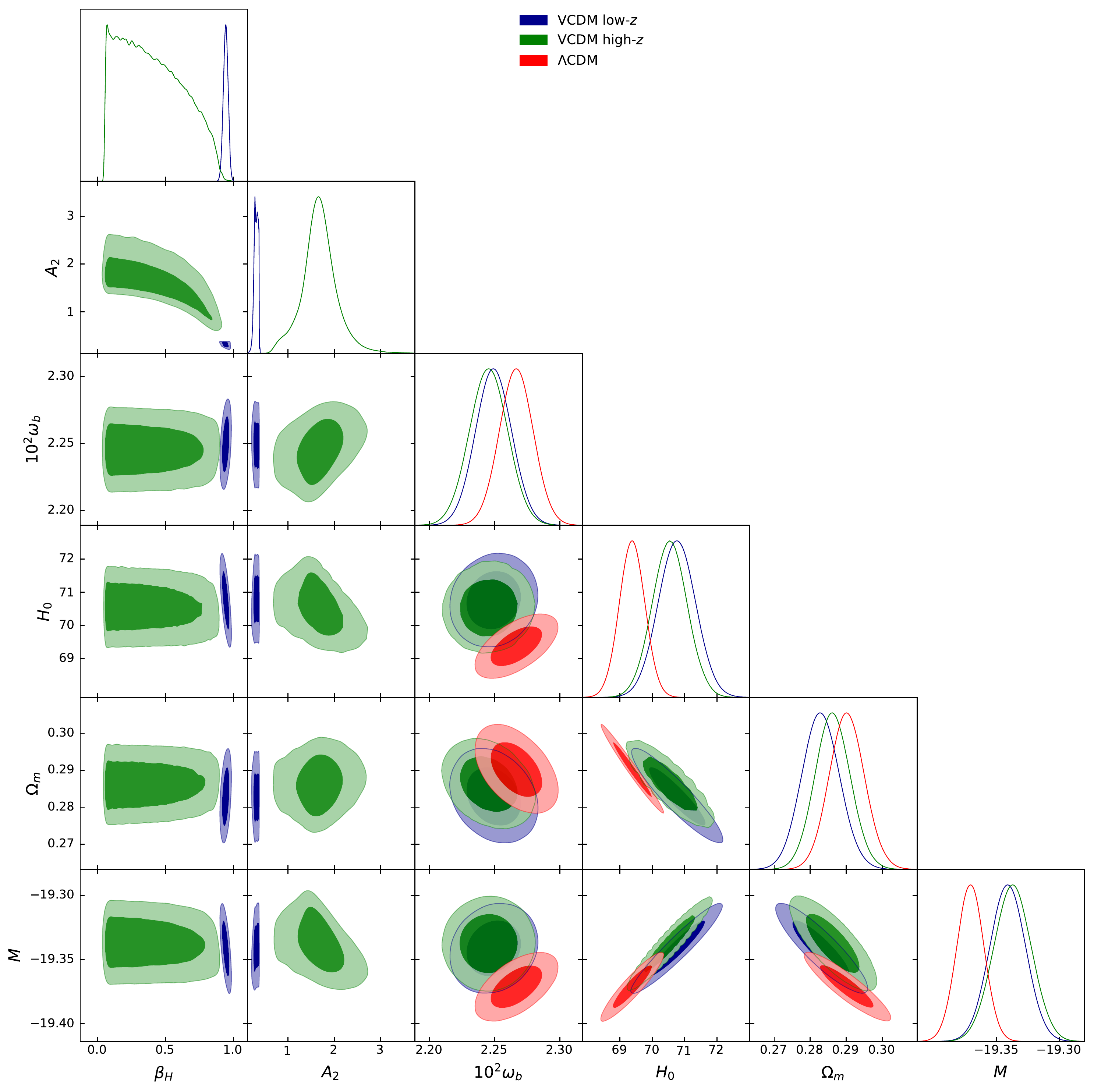}\caption{2-dimensional marginalised likelihoods for the VCDM low$-z$, VCDM high$-z$ and $\Lambda$CDM model fitting against the cosmological data sets.}\label{2dplots}
\end{figure*}

\begin{figure}
\includegraphics[width=0.5\linewidth]{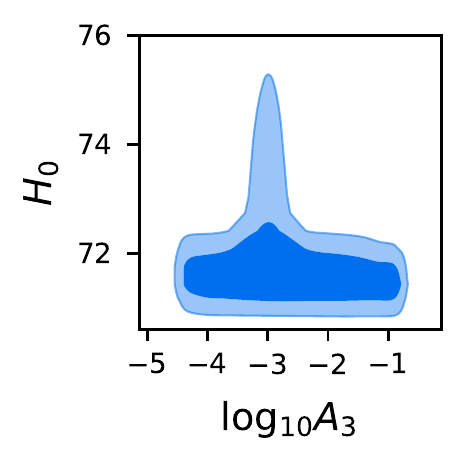}\caption{Degeneracy over the parameter $A_3$.}\label{fig:deg_3p}
\end{figure}

Fig.~\ref{2dplots} shows 2-dimensional marginalised likelihoods for
the cosmological parameters of interest in VCDM model as well as for
$\Lambda$CDM. The Table~\ref{param} gives the values of the parameters
within $2\sigma$'s. It is clear that the parameter $A_2$ has a sharp
upper cut off. This can be understood by the following logic. Both BAO
and Planck data have a better fit for a dynamics for $H(z)$ which
leads to lower values of $H_0$ (compared to local measurements in
$\Lambda$CDM). This behavior still holds for VCDM. But for lower
redshifts (outside the range of BAO-dr12, for which $0.38<z<0.61$),
Pantheon data require larger values for $H(z)$, and the transition
occurs. In order to take into account lower redshift BAO data, we have
also considered small-$z$ BAO data (refer to
table~\ref{chi2_eff-each}). This explains the redshift of transition
(related to the $A_2$ parameter), and probably some similar behavior
will be required if both current and future cosmological data will
keep constraining the $\Lambda$CDM profile for $H(z)$. As for the
width of the transition (related to the $A_3$ parameter, we have fixed
it to the value of $10^{-3}$. This does not lead to a fine-tuning: it
is just a choice. In fact, if we perform a Montecarlo sampling by
adding this third parameter, we find for the same wanted value of
$H_0$ a large degeneracy (about three orders of magnitude at
1$\sigma$) for $A_3$ (see Fig.~\ref{fig:deg_3p}). This only shows that
the data are still not powerful enough to give some insight into this
parameter. For this reason, we have fixed $A_3$ to a reasonable value.

\begin{table*}
\begin{tabular}{l l l l l}
\toprule
& \hspace{1cm}  \textbf{VCDM low$-z$} & \hspace{1cm}  \textbf{VCDM high$-z$}  & \hspace{1cm}  \textbf{$\Lambda$CDM}\\  
\toprule
\textbf{Parameters} & \hspace{1cm}  \textbf{$95\%$ limits}& \hspace{1cm} \textbf{$95\%$ limits} & \hspace{1cm} \textbf{$95\%$ limits}  \\
\midrule
\vspace{0.1cm} 
$\beta_{H }$ & \hspace{1cm} $0.947^{+0.031}_{-0.037}$ & \hspace{1cm} $\leq 0.80$ & \hspace{1cm} $-$ \\
\vspace{0.1cm} 
$A_2$ & \hspace{1cm} $0.295^{+0.086}_{-0.052}$& \hspace{1cm} $1.82^{+0.69}_{-0.93}$  & \hspace{1cm} $-$ \\
\vspace{0.1cm} 
 $10^{2}\omega_{\rm b }$ & \hspace{1cm} $2.254^{+0.022}_{-0.032}$ & \hspace{1cm} $2.240^{+0.034}_{-0.024}$  & \hspace{1cm} $2.270^{+0.023}_{-0.029}$ \\
\vspace{0.1cm} 
$\tau_{\rm reio }$ & \hspace{1cm} $0.054^{+0.018}_{-0.013}$  & \hspace{1cm} $0.053^{+0.017}_{-0.014}$ & \hspace{1cm} $0.061^{+0.015}_{-0.017}$ \\
\vspace{0.1cm} 
  $n_{s }$ & \hspace{1cm} $0.9677^{+0.0771}_{-0.0769}$ & \hspace{1cm} $0.9664^{+0.008}_{-0.009}$ & \hspace{1cm} $0.9736^{+0.0066}_{-0.0076}$\\
  \vspace{0.1cm} 
$10^{10}A_{s }$ & \hspace{1cm} $3.043^{+0.036}_{-0.028}$& \hspace{1cm} $3.044^{+0.032}_{-0.032}$ & \hspace{1cm} $3.052^{+0.031}_{-0.036}$\\  
  \vspace{0.1cm} 
$H_{0 }$ & \hspace{1cm} $70.83^{+1.07}_{-1.13}$ & \hspace{1cm} $70.49^{+1.11}_{-1.09}$ & \hspace{1cm} $69.40^{+0.76}_{-0.8}$\\
\vspace{0.1cm} 
$\Omega_{\rm m }$ & \hspace{1cm} $0.282^{+0.011}_{-0.009}$ & \hspace{1cm} $0.2865^{+0.0096}_{-0.0097}$ & \hspace{1cm} $0.2899^{+0.0101}_{-0.0092}$\\
$M$ & \hspace{1cm} $-19.34^{+0.03}_{-0.03}$ & \hspace{1cm} $-19.34^{+0.03}_{-0.03}$ & \hspace{1cm} $-19.37^{+0.02}_{-0.02}$\\

\bottomrule
\end{tabular}
\caption{One-dimensional 2$\sigma$ constraints for the cosmological parameters of interest after the estimation with the cosmological data sets considered.}\label{param}
\end{table*}

\begin{figure}
 \includegraphics[width=0.8\linewidth]{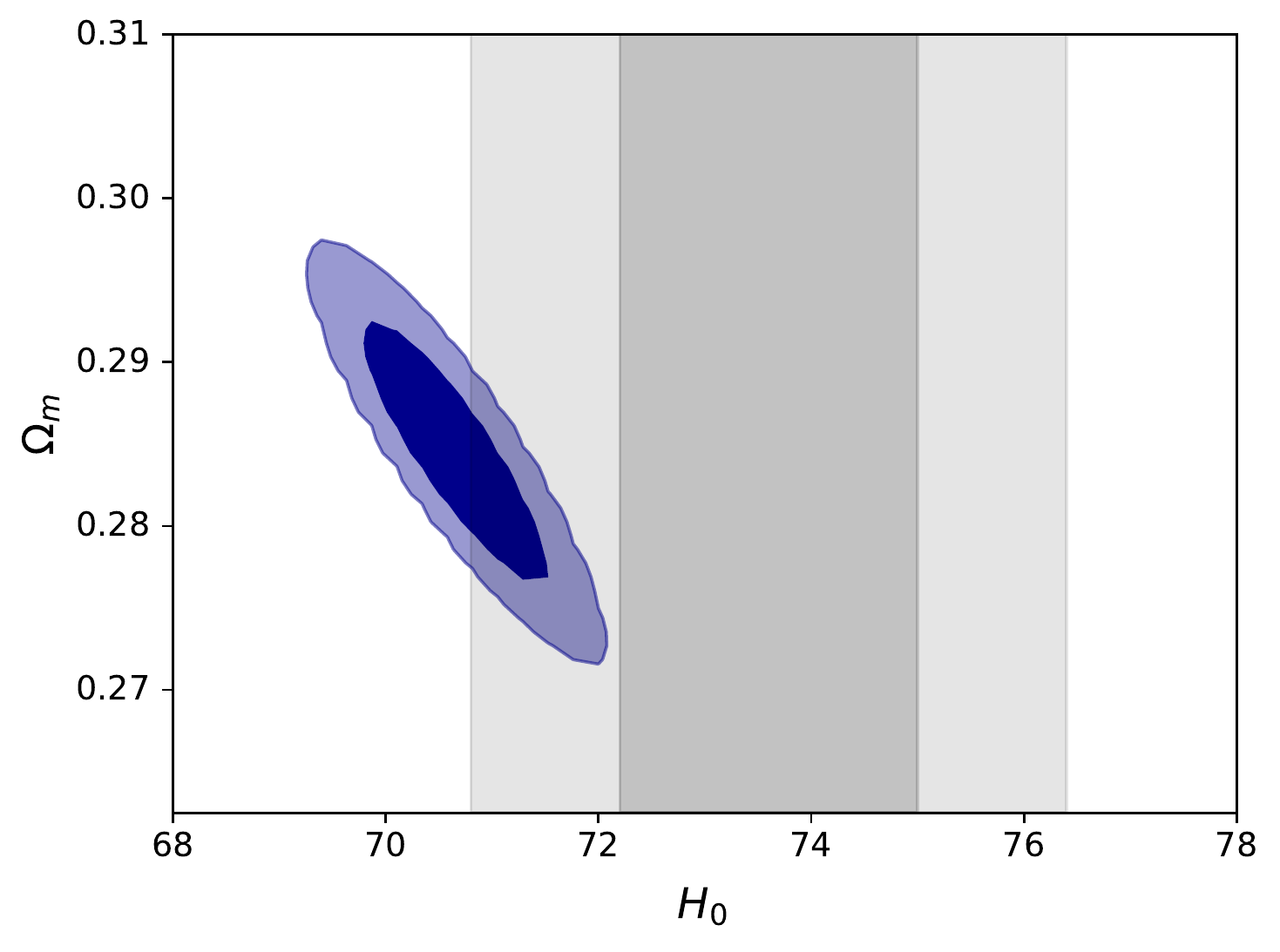}
 \caption{$\Omega_{\rm m}-H_0$ contour showing the $H_0-$tension is reduced compared with the local determination of $H^{\rm{loc}}_0$ within the VCDM theory. The black shade is $H^{\rm{loc}}_0$.}\label{om_H0}
\end{figure}

From Table~\ref{param}, it is interesting to notice that the value of
$\beta_H$ does not reach 1 even at 2$\sigma$. It means that the data
prefer VCDM over $\Lambda$CDM. We find that the bestfit value of
Hubble expansion rate today is $H_0=70.83$ and $H_0=70,49$ for low$-z$
and high$-z$ respectively, which indicates that the tension is
reduced. However we need to determine the local value of the Hubble
expansion rate today $H^{\rm{loc}}_{0}$ following the analysis
explained in~\cite{Camarena:2019moy,Benevento:2020fev} (refer to
Appendix~\ref{h0loc} for details). We find that the local value of the
Hubble expansion today is $H^{\rm{loc}}_{0}=73.6 \pm
1.4$. Fig.~\ref{om_H0} shows the contour of $\Omega_m$ and $H_0$
determined by the MCMC analysis and also the $H^{\rm{loc}}_{0}$ with
$2\sigma$ error bars.

\begin{figure}
\includegraphics[width=1\linewidth]{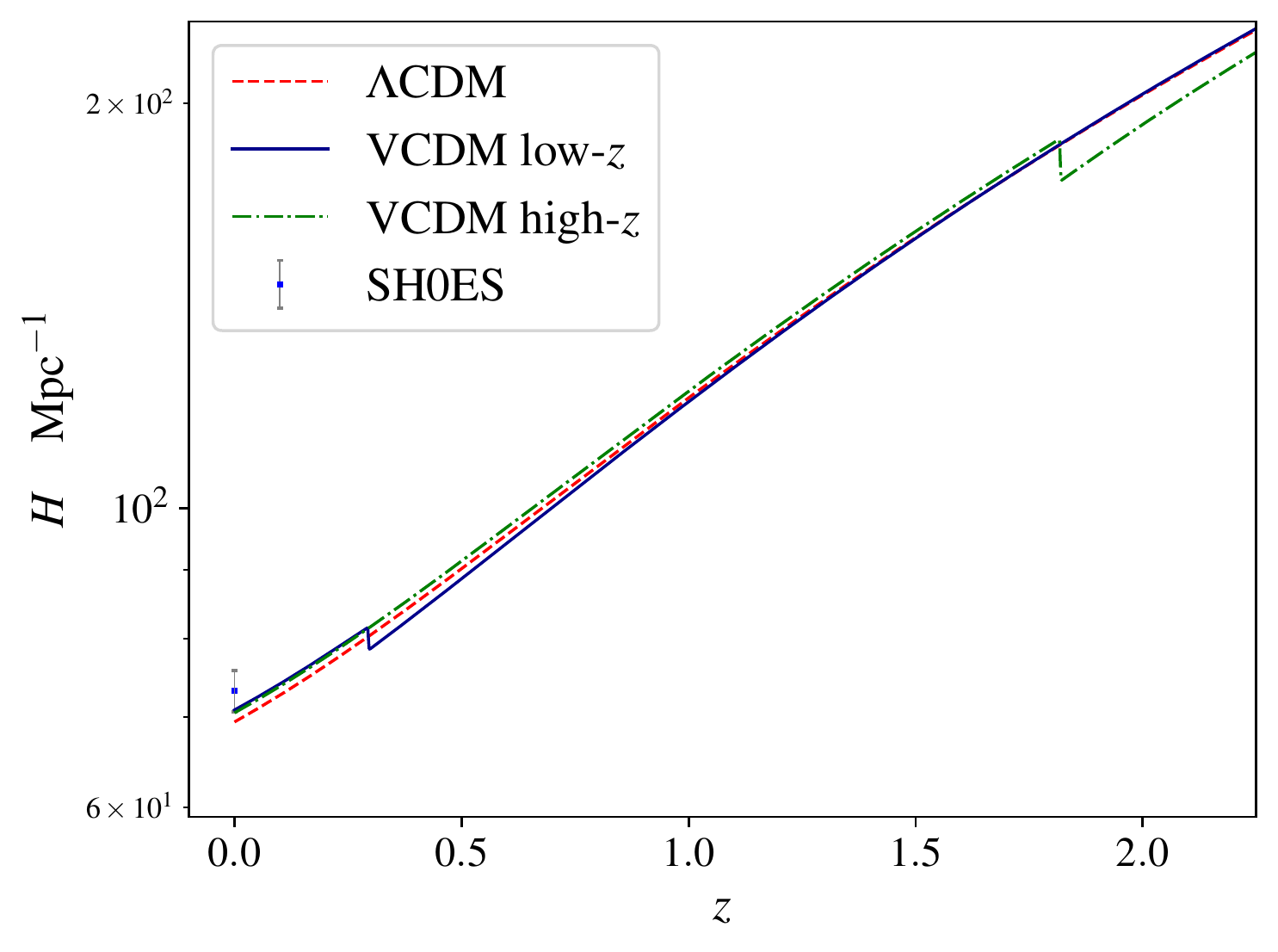}
\caption{Zoomed version of $H$ vs $z$ plot. Here we can see two independent transitions in the $H(z)$ at very low redshift around $z \simeq 0.3$ and $z \simeq 1.8$.}\label{H_zoom}
\end{figure}

Let us look at the behavior of $H(z)$ in VCDM. Fig.\ref{H_zoom}, shows
the two independent transitions at $z \simeq 3$ and $z \simeq 1.8$
respectively.  As explained earlier, an intuitive picture from
equation (\ref{eq:Fried}) gives the parameter $\beta_{H}$ as the
amplitude of the transition, $A_2$ the location of the redshift $z$ at
which the transition happens and $A_{3}$ is the width of such
transition. It is clear from the choice of $H(z)$ that this is a
low-redshift resolution for Hubble tension. Similar proposals have
been made in~\cite{Keeley:2019esp,Vattis:2019efj, Choi:2019jck,
  DeFelice:2020sdq, Hryczuk:2020jhi}. Refer
to~\cite{DiValentino:2021izs} for a review of possible solutions to
Hubble tension.

In this report we showed that the notorious $H_{0}$ tension can be
addressed by a very minimal modification to the standard cosmological
model, dubbed VCDM. We see that the value of $H^{\rm {loc}}_0$
estimated with this theory reduces the $H_0$ tension. It is also
noticed that this model fits the cosmological data sets better than
$\Lambda$CDM. On using Planck-Pantheon-BAO data alone we have seen two
minima existing (which mostly differ by the value of redshift at which
the transition in $H(z)$ occurs). On adding the three most recent
measurements of $H_0$ (SH0ES, H0LiCOW and MEGAMASER) to the above
chosen data sets, the two minima still exist with very similar values
of $\chi^2$ for each of them. These two minima not only differ, as
already stated, by the value of the redshift where the $H(z)$
transition occurs, but also by the value of $\beta_H$ (corresponding
to the amplitude of the transition), but in any case $\beta_H<1$ at
95\% confidence level ($\beta_H=1$ would make VCDM exactly coincide
with $\Lambda$CDM). We find indeed that, compared to $\Lambda$CDM, the
total fitness parameter is improved by $\Delta \chi^2 \simeq 12$.
Background and perturbation variables are stable and finite. Hence we
propose the VCDM model as a possible solution to the $H_0$ tension
(giving the huge freedom for the choice of any wanted $H(z)>0$, still
compatible with a background evolution with no-ghosts, gradient
instabilities, or extra fifth-force gravitational degrees of
freedom). We need further investigation to look at whether VCDM can
address the tension in the large-scale structure, $S_{8}$ along with
the $H_{0}$ tension, although as already shown in
\cite{DeFelice:2020eju}, for this theory $G_{\rm eff}/G_N=1$, at short
scales. The behaviour of reducing $H_{0}$ tension within this theory
sounds promising and it would be interesting to test this behaviour
with future cosmology surveys like EUCLID~\cite{Amendola:2012ys} and
LSST~\cite{Abate:2012za}. Finally we want to stress here that
violations of 4D diffeo are only present in the gravity sector at
cosmological scales. Since gravity is modified in the IR limit, and
because of the absence of any extra gravitational mode other than the
standard tensorial gravitons, we expect that graviton loop corrections
to be negligible. Therefore matter sector Lagrangians are fully
covariant in 4D, up to $\Mpl^2$-suppressed, tiny radiative
corrections.

\appendix
\section{Calculation of $H_0^{\rm {loc}}$}\label{h0loc}
Here we explain how to determine the local value of Hubble expansion
rate today for a given theoretical model. We follow the method
explained in~\cite{Camarena:2019moy,Benevento:2020fev}.

The apparent magnitude of a supernovae at a redshift $z$ is given by
\begin{equation}
m^t_B(z)=5 \log_{10}\left[\frac{d_L(z)}{1 \rm{Mpc}}\right]+25+M_B\,,
\end{equation}
where $d_L(z)$ is the luminosity distance and $M_B$ is the absolute
magnitude. The superscript $t$ in $m^t_B(z)$ stands
  for theoretical apparent magnitude. The luminosity distance is
given by
\begin{equation}
d_{L}=\frac{c}{H_{0}}(1+z)\int_{0}^{z}\frac{dz'}{E(z')}\,,
\end{equation}
where $E=H/H_{0}$. Then the apparent magnitude can be rewritten as
\begin{eqnarray}
  m_{B}^{t} & = & 5\log_{10}\!\left[(1+z)\int_{0}^{z}\frac{dz'}{E(z')}\right]\\ \nonumber
  & &-5\log_{10}\!\left[\frac{(1{\rm Mpc})H_{0}}{c}\right] +25+M_{B}\,.
\end{eqnarray}
Now we define
\begin{eqnarray}
\tilde{m}_{B}^{t} & = & m_{B}^{t}-M_{B}+5\log_{10}\!\left[\frac{(1{\rm Mpc})H_{0}}{c}\right]\nonumber \\
 & = & 5\log_{10}\!\left[(1+z)\int_{0}^{z}\frac{dz'}{E(z')}\right]+25\,,
\end{eqnarray}
which does not depend on both $H_{0}$ and $M_{B}$, but only on the
dynamics of $E(z)$, which is a function of the other parameters of VCDM. We would then introduce the residual
\[
m_{B,i}-m_{B,i}^{t}=m_{B,i}-\tilde{m}_{B,i}^{t}-M_{B}+5\log_{10}\!\left[\frac{(1{\rm Mpc})H_{0}}{c}\right],
\]
to find a $\chi^{2}$ distribution out of it. On calling
\begin{eqnarray}
  W_{i}& =& m_{B,i}-\tilde{m}_{B,i}^{t} \\ \nonumber
  & =& m_{B,i}-\left(5\log_{10}\!\left[(1+z)\int_{0}^{z}\frac{dz'}{E(z')}\right]+25\right),
\end{eqnarray}
we then need to find the residues on the variable
\begin{eqnarray}
\chi^{2} & = & (m_{B,i}-m_{B,i}^{t})\Sigma_{ij}^{-1}(m_{B,j}-m_{B,j}^{t})\nonumber \\
         & = & \left(W_{i}-M_{B}+5\log_{10}\!\left[\frac{(1{\rm Mpc})H_{0}}{c}\right]\right)\Sigma_{ij}^{-1}\\ \nonumber
  & & \left(W_{j}-M_{B}+5\log_{10}\!\left[\frac{(1{\rm Mpc})H_{0}}{c}\right]\right).
\end{eqnarray}

Now consider
\begin{equation}
\bar{d}_{L}\equiv(1+z)\int_{0}^{z}\frac{dz'}{E(z')}\,,
\end{equation}
so that
\begin{equation}
\bar{d}'_{L}  = \frac{dz}{dN} \frac{\bar{d}_{L}}{dz} = \bar{d}_{L}+\frac{(1+z)^{2}}{E(z)}\,,\\
\end{equation}
where we have used
\begin{equation}
\frac{dz}{dN}=1+z\,,
\end{equation}
considering  $N=\ln(a_{0}/a)=\ln(1+z)$.
Now we can solve for $\bar{d}_{L}(z)$, given the initial conditions
$\bar{d}_{L}(0)=0=z(0)$.

Once we have the quantities $\bar{d}_{L}$ for any data-redshift,
we have $W_{i}$ so that we are able to find
\begin{eqnarray}
S_{0} & \equiv & V^{T}\Sigma^{-1}V\,,\\
S_{1} & \equiv & W^{T}\Sigma^{-1}V\,,
\end{eqnarray}
where $V_{i}=1$ and $\Sigma_{ij}$ is the covariance matrix.

Finally, the mean value and the variance of $H^{\rm {loc}}_0$ can be determined by the log-normal distribution
\begin{eqnarray}
  H^{\rm{loc}}_0 &=& e^{{\mu}_{\ln}+\frac{1}{2}\sigma^2_{\ln}}\,, \\
  \sigma^2_{H^{\rm{loc}}_0} & = & \left( e^{\sigma^2_{\ln}}-1\right)e^{2\mu_{\ln}+\sigma^2_{\ln}}\,,
\end{eqnarray}
where
\begin{eqnarray}
  \mu_{\ln} &=& \frac{\ln 10}{5} \left[\bar{M}_B+\frac{\ln 10}{5} \left(\sigma^2_M + \frac{1}{S_0}\right) - \frac{S_1}{S_0}\right], \\
  \sigma_{\ln} &=& \frac{\ln 10}{5}\sqrt{\sigma^2_M+\frac{1}{S_0}}\,,
\end{eqnarray}
which, in turn, only depends on $M_{B}$, $\sigma_{M}^{2}$, $S_{0}$ and
$S_{1}$. Also one should notice that $S_{0}$ is only given by the
data, and it does not depend on the model, but $S_{1}$ does depend on
the values of $\bar{d}_{L}$'s, and this will affect
$H_{0}^{{\rm loc}}$ for different models. To get a log-normal
distribution we assumed a Gaussian distribution for $M_{B}$ and have
marginalized over it (refer to~\cite{Camarena:2019moy} for details).

Now we select the the supernovae data and the covariance matrix from
Pantheon\footnote{https://github.com/dscolnic/Pantheon} dataset up to
$z\leq 0.15$, according to~\cite{Camarena:2019moy}. Then we integrate
the luminosity distance with respect to $N$ to find
$S_1$. From~\cite{Camarena:2019moy} we use the values
$\bar{M}_B=-19.2322$ and $\sigma_M=0.0404$. Hence we find the values
of $H^{\rm{loc}}_0$ and $\sigma^2_{H^{\rm{loc}}_0}$.

Further study is necessary to understand whether this late time change
can fully address today's cosmological puzzles, including the $S_8$
tension. However, we think this study might help people focusing
their efforts on finding the best profile for $H(z)$ which can model the
data sets.

\begin{acknowledgments}
We thank George Efstathiou for useful comments. The work of A.D.F.\ was supported by Japan Society for
the Promotion of Science Grants-in-Aid for Scientific Research No.\ 20K03969.
 The work of S.M.\ was supported by JSPS KAKENHI Grant Numbers 
 17H02890, 17H06359, and by WPI MEXT, Japan. 
  M.C.P.\ acknowledges the support
 from the Japanese Government (MEXT) scholarship for Research
 Student. Numerical computation in this work was carried out at the
 Yukawa Institute Computer Facility. 
\end{acknowledgments}

\bibliographystyle{unsrt}
\bibliography{bibliography}

\begin{thebibliography}{10}

\bibitem{Riess:2019cxk}
Adam~G. Riess, Stefano Casertano, Wenlong Yuan, Lucas~M. Macri, and Dan
  Scolnic.
\newblock {Large Magellanic Cloud Cepheid Standards Provide a 1\% Foundation
  for the Determination of the Hubble Constant and Stronger Evidence for
  Physics beyond $\Lambda$CDM}.
\newblock {\em Astrophys. J.}, 876(1):85, 2019.

\bibitem{Wong:2019kwg}
Kenneth~C. Wong et~al.
\newblock {H0LiCOW XIII. A 2.4\% measurement of $H_{0}$ from lensed quasars:
  $5.3\sigma$ tension between early and late-Universe probes}.
\newblock 7 2019.

\bibitem{Reid:2008nm}
M.J. Reid, J.A. Braatz, J.J. Condon, L.J. Greenhill, C.~Henkel, and K.Y. Lo.
\newblock {The Megamaser Cosmology Project: I. VLBI observations of UGC 3789}.
\newblock {\em Astrophys. J.}, 695:287--291, 2009.

\bibitem{Freedman:2019jwv}
Wendy~L. Freedman et~al.
\newblock {The Carnegie-Chicago Hubble Program. VIII. An Independent
  Determination of the Hubble Constant Based on the Tip of the Red Giant
  Branch}.
\newblock 7 2019.

\bibitem{Aghanim:2018eyx}
N.~Aghanim et~al.
\newblock {Planck 2018 results. VI. Cosmological parameters}.
\newblock 7 2018.

\bibitem{Bernal:2016gxb}
Jose~Luis Bernal, Licia Verde, and Adam~G. Riess.
\newblock {The trouble with $H_0$}.
\newblock {\em JCAP}, 10:019, 2016.

\bibitem{Riess:2020sih}
Adam~G. Riess.
\newblock {The Expansion of the Universe is Faster than Expected}.
\newblock {\em Nature Rev. Phys.}, 2(1):10--12, 2019.

\bibitem{DeFelice:2020eju}
Antonio De~Felice, Andreas Doll, and Shinji Mukohyama.
\newblock {A theory of type-II minimally modified gravity}.
\newblock {\em JCAP}, 09:034, 4 2020.

\bibitem{Liddle:2007fy}
Andrew~R Liddle.
\newblock {Information criteria for astrophysical model selection}.
\newblock {\em Mon. Not. Roy. Astron. Soc.}, 377:L74--L78, 2007.

\bibitem{Aghanim:2019ame}
N.~Aghanim et~al.
\newblock {Planck 2018 results. V. CMB power spectra and likelihoods}.
\newblock 2019.

\bibitem{Beutler:2011hx}
Florian Beutler, Chris Blake, Matthew Colless, D.~Heath Jones, Lister
  Staveley-Smith, Lachlan Campbell, Quentin Parker, Will Saunders, and Fred
  Watson.
\newblock {The 6dF Galaxy Survey: Baryon Acoustic Oscillations and the Local
  Hubble Constant}.
\newblock {\em Mon. Not. Roy. Astron. Soc.}, 416:3017--3032, 2011.

\bibitem{Ross:2014qpa}
Ashley~J. Ross, Lado Samushia, Cullan Howlett, Will~J. Percival, Angela Burden,
  and Marc Manera.
\newblock {The clustering of the SDSS DR7 main Galaxy sample -- I. A 4 per cent
  distance measure at $z = 0.15$}.
\newblock {\em Mon. Not. Roy. Astron. Soc.}, 449(1):835--847, 2015.

\bibitem{Alam:2016hwk}
Shadab Alam et~al.
\newblock {The clustering of galaxies in the completed SDSS-III Baryon
  Oscillation Spectroscopic Survey: cosmological analysis of the DR12 galaxy
  sample}.
\newblock {\em Mon. Not. Roy. Astron. Soc.}, 470(3):2617--2652, 2017.

\bibitem{Scolnic:2017caz}
D.M. Scolnic et~al.
\newblock {The Complete Light-curve Sample of Spectroscopically Confirmed SNe
  Ia from Pan-STARRS1 and Cosmological Constraints from the Combined Pantheon
  Sample}.
\newblock {\em Astrophys. J.}, 859(2):101, 2018.

\bibitem{Riess:2020fzl}
Adam~G. Riess, Stefano Casertano, Wenlong Yuan, J.~Bradley Bowers, Lucas Macri,
  Joel~C. Zinn, and Dan Scolnic.
\newblock {Cosmic Distances Calibrated to 1\% Precision with Gaia EDR3
  Parallaxes and Hubble Space Telescope Photometry of 75 Milky Way Cepheids
  Confirm Tension with $\Lambda$CDM}.
\newblock {\em Astrophys. J. Lett.}, 908(1):L6, 2021.

\bibitem{Camarena:2019moy}
David Camarena and Valerio Marra.
\newblock {Local determination of the Hubble constant and the deceleration
  parameter}.
\newblock {\em Phys. Rev. Res.}, 2(1):013028, 2020.

\bibitem{Benevento:2020fev}
Giampaolo Benevento, Wayne Hu, and Marco Raveri.
\newblock {Can Late Dark Energy Transitions Raise the Hubble constant?}
\newblock {\em Phys. Rev. D}, 101(10):103517, 2020.

\bibitem{Camarena:2021jlr}
David Camarena and Valerio Marra.
\newblock {On the use of the local prior on the absolute magnitude of Type Ia
  supernovae in cosmological inference}.
\newblock 1 2021.

\bibitem{Efstathiou:2021ocp}
George Efstathiou.
\newblock {To H0 or not to H0?}
\newblock 3 2021.

\bibitem{Aghanim:2016sns}
N.~Aghanim et~al.
\newblock {Planck intermediate results. LI. Features in the cosmic microwave
  background temperature power spectrum and shifts in cosmological parameters}.
\newblock {\em Astron. Astrophys.}, 607:A95, 2017.

\bibitem{Motloch:2019gux}
Pavel Motloch and Wayne Hu.
\newblock {Lensinglike tensions in the $Planck$ legacy release}.
\newblock {\em Phys. Rev. D}, 101(8):083515, 2020.

\bibitem{Blas:2011rf}
Diego Blas, Julien Lesgourgues, and Thomas Tram.
\newblock {The Cosmic Linear Anisotropy Solving System (CLASS) II:
  Approximation schemes}.
\newblock {\em JCAP}, 1107:034, 2011.

\bibitem{Pookkillath:2019nkn}
Masroor~C. Pookkillath, Antonio De~Felice, and Shinji Mukohyama.
\newblock {Baryon Physics and Tight Coupling Approximation in Boltzmann Codes}.
\newblock {\em Universe}, 6:6, 2020.

\bibitem{Audren:2012wb}
Benjamin Audren, Julien Lesgourgues, Karim Benabed, and Simon Prunet.
\newblock {Conservative Constraints on Early Cosmology: an illustration of the
  Monte Python cosmological parameter inference code}.
\newblock {\em JCAP}, 1302:001, 2013.

\bibitem{Brinckmann:2018cvx}
Thejs Brinckmann and Julien Lesgourgues.
\newblock {MontePython 3: boosted MCMC sampler and other features}.
\newblock {\em Phys. Dark Univ.}, 24:100260, 2019.

\bibitem{Lewis:2019xzd}
Antony Lewis.
\newblock {GetDist: a Python package for analysing Monte Carlo samples}.
\newblock 10 2019.

\bibitem{Keeley:2019esp}
Ryan~E. Keeley, Shahab Joudaki, Manoj Kaplinghat, and David Kirkby.
\newblock {Implications of a transition in the dark energy equation of state
  for the $H_0$ and $\sigma_8$ tensions}.
\newblock {\em JCAP}, 12:035, 2019.

\bibitem{Vattis:2019efj}
Kyriakos Vattis, Savvas~M. Koushiappas, and Abraham Loeb.
\newblock {Dark matter decaying in the late Universe can relieve the H0
  tension}.
\newblock {\em Phys. Rev. D}, 99(12):121302, 2019.

\bibitem{Choi:2019jck}
Gongjun Choi, Motoo Suzuki, and Tsutomu~T. Yanagida.
\newblock {Quintessence Axion Dark Energy and a Solution to the Hubble
  Tension}.
\newblock {\em Phys. Lett. B}, 805:135408, 2020.

\bibitem{DeFelice:2020sdq}
Antonio De~Felice, Chao-Qiang Geng, Masroor~C. Pookkillath, and Lu~Yin.
\newblock {Reducing the $H_{0}$ tension with generalized Proca theory}.
\newblock {\em JCAP}, 08:038, 2020.

\bibitem{Hryczuk:2020jhi}
Andrzej Hryczuk and Krzysztof Jod\l{}owski.
\newblock {Self-interacting dark matter from late decays and the $H_0$
  tension}.
\newblock {\em Phys. Rev. D}, 102(4):043024, 2020.

\bibitem{DiValentino:2021izs}
Eleonora Di~Valentino, Olga Mena, Supriya Pan, Luca Visinelli, Weiqiang Yang,
  Alessandro Melchiorri, David~F. Mota, Adam~G. Riess, and Joseph Silk.
\newblock {In the Realm of the Hubble tension $-$ a Review of Solutions}.
\newblock 3 2021.

\bibitem{Amendola:2012ys}
Luca Amendola et~al.
\newblock {Cosmology and fundamental physics with the Euclid satellite}.
\newblock {\em Living Rev. Rel.}, 16:6, 2013.

\bibitem{Abate:2012za}
Alexandra Abate et~al.
\newblock {Large Synoptic Survey Telescope: Dark Energy Science Collaboration}.
\newblock 11 2012.

\end{thebibliography}

\end{document}